\documentclass[aps,preprint,superscriptaddress,showpacs]{revtex4}
\usepackage{graphicx}
\begin{document}
\bibliographystyle{apsrev}
 
\title{Ground state structure and conductivity of quantum
       wires of infinite length and finite width}

\author{F. Malet}
\affiliation{Departament ECM, Facultat de F\'{\i}sica
Universitat de Barcelona. Diagonal 647
08028 Barcelona, Spain}

\author{M. Pi}
\affiliation{Departament ECM, Facultat de F\'{\i}sica
Universitat de Barcelona. Diagonal 647
08028 Barcelona, Spain}

\author{M. Barranco}
\affiliation{Departament ECM, Facultat de F\'{\i}sica
Universitat de Barcelona. Diagonal 647
08028 Barcelona, Spain}

\author{E. Lipparini}
\affiliation{Dipartimento di Fisica, Universit\`a di Trento,
and INFN sezione di Trento, I-38050 Povo, Italy}

\date{\today}

\begin{abstract} 
We have studied the ground state structure of  quantum strips
within the  local spin-density approximation,
for a range of electronic densities between $\sim$ 5$\times$10$^4$ and
2$\times$10$^6$ cm$^{-1}$ and several strengths of the 
lateral confining potential.
The results have been used to address the conductance $G$ of quantum
strips. At low density, when only one subband is occupied,
the system is fully polarized and $G$  takes a value
which
is close  to 0.7(2e$^2/h$), decreasing with increasing electron
density in  agreement with experiments. 
At higher densities the system becomes paramagnetic and $G$
takes a value near (2e$^2/h$), showing a similar decreasing behaviour
with increasing electron density.
 In both cases, the physical
parameter that determines the value of the conductance is
the ratio $K/K_0$
of the compressibility of the system over the free one. 
\end{abstract}


\maketitle

\section{Introduction}

The conductance of quantum wires, i.e., the linear conductivity per 
unit length,  is a subject of current 
interest since the progress in nanostructure technology
has allowed the fabrication of quasi one-dimensional (1D)
structures. An interesting phenomenon is the observation
of quantization of conductance $G$
in units of $2 e^2/h$, which
reflects the number of active channels
in the transport measurement \cite{van88,Wha88,Kaw89}.
This result is usually explained by considering the allowed
energy subbands of a non interacting 1D electron gas,
where the factor of 2 is due to spin degeneracy.
Interaction effects have been considered for some time,
especially in the framework of the 1D Tomonaga-Luttinger model,
where it is predicted that the conductance is renormalized 
to $G=\gamma (2e^2/h)$, with a parameter $\gamma>1$ for attractive
interactions, $\gamma<1$ for repulsive interactions, and $\gamma=1$
for non interacting electron gas \cite{Ap82,Kan92a,Oga94}.
However, it has been argued \cite{Mas95,Saf95,Pon95,Kaw95,Ore96} that
$\gamma$ should be unity, since the measured conductance is
determined by the non interacting electrons which are injected in the 
wire. 

The conductivity measured in the experiment is not the 
response to the external applied electric field, but to that
which results adding to the applied field that caused by
the induced charge \cite{Izu61,Pin66}, usually  called induced
polarization field. In
other words, the measured linear conductivity is
the screened one, which in the random phase approximation (RPA) used in
previous calculations coincides with the free linear conductivity.
Recent  measurements \cite{Rei01}
in ultra low-disorder quantum wires of finite length exhibit a
conductance structure close to $G=0.7 (2e^2/h)$ evolving continuosly 
with increasing electron density to $G=0.5 (2e^2/h)$, the value expected 
for an ideal spin-split subband. The structure at $G=0.7 (2e^2/h)$
was already observed in low-disorder quantum point contacts (quantum
wires of zero length) by several authors
\cite{Wee91,Tho96,Tho98,Kan98,Kri98,Tsc96}.
In particular, it has been argued \cite{Tho98} that this
structure is a manifestation of electron-electron interactions involving
spin. Theoretically, the structure is interpreted as some form of
spontaneous spin polarization of the system  mediated through the exchange 
interaction \cite{Gold96,Wan96,Ram05,Ber02,Sta03,Hav04}.
The appearance of that structure has been also interpreted as a
manifestation of a Kondo effect in quantum point contacts
\cite{Cro02,Mei02,Hir03}.
Other interpretations of the plateau of $G$ at $0.5 (2e^2/h)$
are based on a possible Wigner crystalization of the 1D
electron gas at low density \cite{Mat04}.
The evolution of the structure with density in
finite length wires is not well understood. For finite length
wires, Reilly et al. \cite{Rei01}
find additional structures in higher subbands which suggest 
that many-body effects are enhanced in longer 1D wires.
Long quantum wires are ideal systems to address one-dimensional
electron transport considering electron-electron interactions
and also in the presence of impurities,
which are relevant to study the Tomonaga-Luttinger liquid
\cite{Oga94,Kan92,Fur93}.

 In this paper we present a 
self-consistent calculation, within local
spin-density approximation (LSDA),  of the ground state of quantum wires
of infinite length and  finite width, and for this reason we prefer
to call them  quantum strips. Their extension in the third direction is
neglected, as in most theoretical descriptions. We apply it to obtain the
screened conductivity in the framework of linear response theory,
showing that in this approximation the key quantity that determines the
conductance of the strip is the ratio
of the compressibility of the system over the free one, $K/K_0$.
Within LSDA, the screened response and $K$ are determined by the
exchange-correlation interaction.
At low densities, only one subband is occupied, the system is
magnetized and the conductance  reduces
to $G=(e^2/h)\sqrt{K/K_0}$, where $K$ is the
compressibility of the spin polarized state, yielding values 
for $G$ which range from $G\simeq0.7 (2e^2/h)$ to $G\simeq0.5 (2e^2/h)$ 
depending on the 
electron density. At larger densities, two degenerated subbands are
occupied, the strip is  paramagnetic and the conductance  is
given by  $G=(2e^2/h)\sqrt{K/K_0}$, where
$K$ is now the compressibility of the paramagnetic state,
 yielding values above -but
close-  to $G= (2e^2/h)$ and  slightly density depending
in the region where the two subbands configuration is
expected to be the physically realized stable phase of the 
strip \cite{Rei99}.

The system is partially magnetized  when
three subbands are occupied, and a paramagnetic state is reached again
at higher density when four subbands are occupied, and so on.
 Our method
allows to calculate the compressibility in all these situations
-although only the cases of one and two occupied subbands are
presented-
and yields values for $G$ that are in qualitative agreement with the
ones found in long wires \cite{Rei01}.
We note that a scattering matrix approach
has been recently employed to calculate the conductance through
a semi-extended barrier or well in the wire \cite{Gud04}.
In this case, $G$ can be
expressed in terms of the incident electron energy $E$ in the form
$G={2 e^2\over h}\sum_n T_n(E)$, where $T_n(E)$ is
the current transmission coefficient for an electron incident in the
nth subband. In our approach, $G$ is calculated in an alternative way
in which the conductance is the linear current response of the system
to an external static electric field 
and hence is an intrinsic property of the
strip, like for example its dielectric constant.

The plan of this work is the following.
In Sec. II we describe the model and obtain the phase diagram of the
strip, i.e., the energy and magnetization as a function of the 1D
electron density along the strip, as well as
its free response. In Sec. III we obtain
the conductance by calculating the screened response to an 
applied electric field in the limit of zero intensity (linear response),
showing that it is mainly determined by the compressibility of the
strip. Some concluding remarks are given in Sec. IV, and the
calculation of the density response function of the strip within
time-dependent LSDA is presented in the Appendix.

\section{Model}

We consider a single, infinitely long quantum strip in the $y$ direction
built on a two-dimensional
electron gas (2DEG) by introducing a confining potential along the $x$
direction.
This potential is assumed to be parabolic ${1\over2}m\omega_0^2\, x^2$.
In the LSDA, the single electron wave functions are given by the
solution of the Kohn-Sham (KS) equations
\begin{eqnarray}
\label{eq1}
\left[\rule{0cm}{0.5cm}\right.
-\frac{1}{2} \nabla_x^2 -\frac{1}{2} \nabla_y^2 
+ \frac{1}{2}\omega_0^2 \,x^2
+\int d{\bf r}^{\prime}\frac{\rho({\bf 
r}^{\prime})}{|{\bf r}-{\bf r}^{\prime}|}
\nonumber\\
+v_{xc}({\bf r}) + w_{xc}({\bf r})\eta_{\sigma}
\left.\rule{0cm}{0.5cm}\right]\varphi^{\sigma}_i({\bf r})
=\varepsilon_{i,\sigma}\,\varphi^{\sigma}_i({\bf r})\; ,
\end{eqnarray}
where $i$ stands for the set of quantum numbers, but spin,
that characterize
the two-dimensional (2D) single-particle wave functions, and
the {\it two-dimensional} electronic density of the strip
is 
$$
\rho({\bf r})=\sum_{i,\sigma}\vert\varphi^{\sigma}_i({\bf r})\vert^2
$$
with ${\bf r}\equiv(x,y)$, $\eta_{\sigma}= 1(-1)$ if $\sigma=\uparrow(\downarrow)$,
and
\begin{equation}
\label{eq2}
v_{xc}({\bf r})=\frac{\partial  {\cal E}_{xc}[\rho({\bf r}),m({\bf
r})]}{\partial\rho({\bf r})}~~~~ ,
~~~~ w_{xc}({\bf r})=\frac{\partial 
{\cal E}_{xc}[\rho({\bf r}),m({\bf r})]}{\partial m({\bf r})}
\end{equation}
with $\rho({\bf r})=\rho^{\uparrow}({\bf r})+\rho^{\downarrow}({\bf
r})$ and $m({\bf r})=\rho^{\uparrow}({\bf r})-\rho^{\downarrow}
({\bf r})$.

The exchange-correlation energy per unit surface
${\cal E}_{xc}$ has been constructed from the
results on the nonpolarized
and fully polarized 2DEG \cite{Tan89} in the same way as in
Refs. \cite{Kos97,Ser99}, i.e., using the two-dimensional von Barth
and Hedin \cite{Hed72} prescription to interpolate between both regimes.
In Eq. (\ref{eq1}) we have used effective atomic units
($\hbar=e^2/\epsilon=m=1$),
where $\epsilon$ is the dielectric constant  and 
$m$ is the electron effective mass. In units of the bare electron mass
$m_e$ one has $m=m^*m_e$. In this system of units, the length
unit is the effective Bohr radius $a_0^*=a_0\,\epsilon/m^*$, and the
energy unit is the effective Hartree $H^*=H m^*/\epsilon^2$. For
GaAs we have taken $\epsilon=12.4$ and $m^*=0.067$, which yields
$a_0^*=97.9$ \AA $\,$ and $H^*=11.9$ meV.

Translational invariance along the $y$-direction imposes solutions of
Eq. (\ref{eq1}) of the form
\begin{equation}
\label{eq3}
\varphi^{\sigma}_i({\bf r})={1\over\sqrt{L}}e^{i k y}\phi^{\sigma}_n(x)~,
\end{equation}
where $n=0,1,2, \ldots$ is the subband index.
Inserting Eq. (\ref{eq3}) into Eq. (\ref{eq1}) one gets
\begin{eqnarray}
\label{eq4}
\left[\rule{0cm}{0.5cm}\right.
-\frac{1}{2}\nabla_x^2
+ \frac{1}{2}\omega_0^2 \, x^2
+\int \int d x^{\prime}d y^{\prime}\frac{\rho(
x^{\prime})}{\sqrt{(x-x^{\prime})^2+(y-y^{\prime})^2}}
\nonumber\\
+v_{xc}(x) + w_{xc}(x)\eta_{\sigma}
\left.\rule{0cm}{0.5cm}\right]\phi^{\sigma}_n(x)
=\epsilon_{n,\sigma}\phi^{\sigma}_n(x)\; ,
\end{eqnarray}
where we have introduced the
band-head energy $\epsilon_{n,\sigma}$
\begin{equation}
\label{eq5}
\epsilon_{n,\sigma}=\varepsilon_{i,\sigma}-{k^2\over2}
\end{equation}
and the 2D density, which is $y$ independent, reads
\begin{equation}
\label{eq6}
\rho_{\sigma}(x)={1\over\pi}\sum_n
\sqrt{2(\epsilon_F-\epsilon_{n,\sigma})}\;
\vert\phi_n^{\sigma}(x)\vert^2~.
\end{equation}
The 1D electron density $\rho_1$ is obtained integrating over $x$:
\begin{equation}
\label{eq7}
\rho_{1}=
{1\over\pi}\sum_{n,\sigma}\sqrt{2(\epsilon_F-\epsilon_{n,\sigma})}
= \rho_1^{\uparrow} + \rho_1^{\downarrow} ~.
\end{equation}
The Fermi energy $\epsilon_F$ fixes the number of subbands
that are filled in the ground state of the strip.
For each  value of $\rho_{1}$, it is determined by solving the KS
Eqs. (\ref{eq4})  self-consistently under the condition that
Eq. (\ref{eq7}) is fulfilled.

For an infinite strip like the one considered here,
the Hartree potential
\begin{equation}
\label{eq8a}
V_H=\int \int d x^{\prime}d y^{\prime}\frac{\rho(
x^{\prime})}{\sqrt{(x-x^{\prime})^2+(y-y^{\prime})^2}} 
\end{equation}
is obviously divergent.
As in the case of the homogenous electron gas in two or three dimensions, 
this requires the introduction
of a neutralizing positive background. One possible way is to
assume that this background is such that the Hartree potential is
strictly cancelled out, and only the exchange and correlation 
energy terms
appear in the KS equations \cite{Rei99}. Another possibility is to introduce 
a positive background that only cancels out the divergency in the Hartree
potential.
With this goal in mind, let us write:  
\begin{equation}
\label{eq8}
V_H=\lim_{q\to0}\, \int \int d x^{\prime}d y^{\prime} \,
e^{iq(y-y^{\prime})}\frac{\rho(
x^{\prime})}{\sqrt{(x-x^{\prime})^2+(y-y^{\prime})^2}}
=\lim_{q\to0}\, 2\int d x^{\prime}\rho(x^{\prime})K_0[
q(x-x^{\prime})]~, \end{equation}
where $K_0$ is the zeroth order Bessel function of second kind
\cite{Abr70}. Since for small arguments  $K_0(s)\simeq-ln(s)$, one gets
\begin{equation}
\label{eq9}
V_H=\lim_{q\to0} \, -2\int d x^{\prime}\rho(x^{\prime})\,ln(qa) 
-2\int dx^{\prime}\rho(x^{\prime}) ln\left| {x-x^{\prime}\over
a}\right|~, 
\end{equation}
where $a$ is an arbitrary length that in the following will be chosen equal to
the confinement length $a=\sqrt{1/\omega_0}$, as in Ref. \cite{Gud95}.
The first term of Eq. (\ref{eq9}) is divergent and is cancelled
by a similar term arising from the background charges. The second term
is what we take as effective
1D Hartree potential, which is logarithmic on an arbitrary length
scale
\begin{equation}
\label{eq10}
V_H(x)= -2\int d x^{\prime}\rho(x^{\prime})ln\left| {x-x^{\prime}\over
a}\right|~. 
\end{equation}
This  potential  has been already considered in Hartree
calculations of quantum wires \cite{Gud95,Ste96}, and may account for
possible local charge nonneutrality in systems with a finite width
\cite{Cam96}, like
the strips considered here. In the following we carry out
Kohn-Sham calculations for both extreme  models,
i.e., when the logarithmic Hartree term  Eq. (\ref{eq10}) is
active, and when this term is neglected due to the assumption of a 
complete cancellation of the Hartree potential Eq. (\ref{eq8a}).

The energy per unit length of the strip can be calculated as
\begin{eqnarray}
\label{eq11}
{E\over L}=\sum_{n,\sigma}
\left[{1\over6\pi}
[2(\epsilon_F-\epsilon_{n,\sigma})]^{3/2}
+{1\over\pi}\sqrt{2(\epsilon_F-\epsilon_{n,\sigma})}
\,\epsilon_{n,\sigma}\right]
+\int d x \,{\cal E}_{xc}(x)
\nonumber\\
 - \int dx \,\rho(x) \,v_{xc}(x) - \int dx \,m(x) \,w_{xc}(x)
+\int dx\, dx^{\prime}\,\rho(x)\,\rho(x^{\prime})\,ln\left|
{x-x^{\prime}\over a}\right| \, .
\end{eqnarray}
The energy per electron $E/N$ is obtained dividing Eq. (\ref{eq11})
by $\rho_1$.

\subsection{Phase diagram of the quantum strip in the Kohn-Sham framework}

We have solved the KS Eqs. for strips confined by three different values
of the harmonic potential, namely $\omega_0=2, 4$, and 6 meV, and  for
$\rho_1$ densities up to filling 6 subbands. Values of $\omega_0$
between 2.5 and 3.5 meV have been determined 
for the gate voltage close to the threshold of the second subband 
for long quantum wires using a magnetic depopulation technique 
\cite{Tar95}. We discuss the case  in which
the logarithmic Hartree potential has been included, which we call Hartree model,
and the case in which the Hartree potential is not taken into account in the KS 
Eqs., which we call the exchange-correlation model.

\begin{figure}

\centerline{\includegraphics[height=20cm,angle=-90]{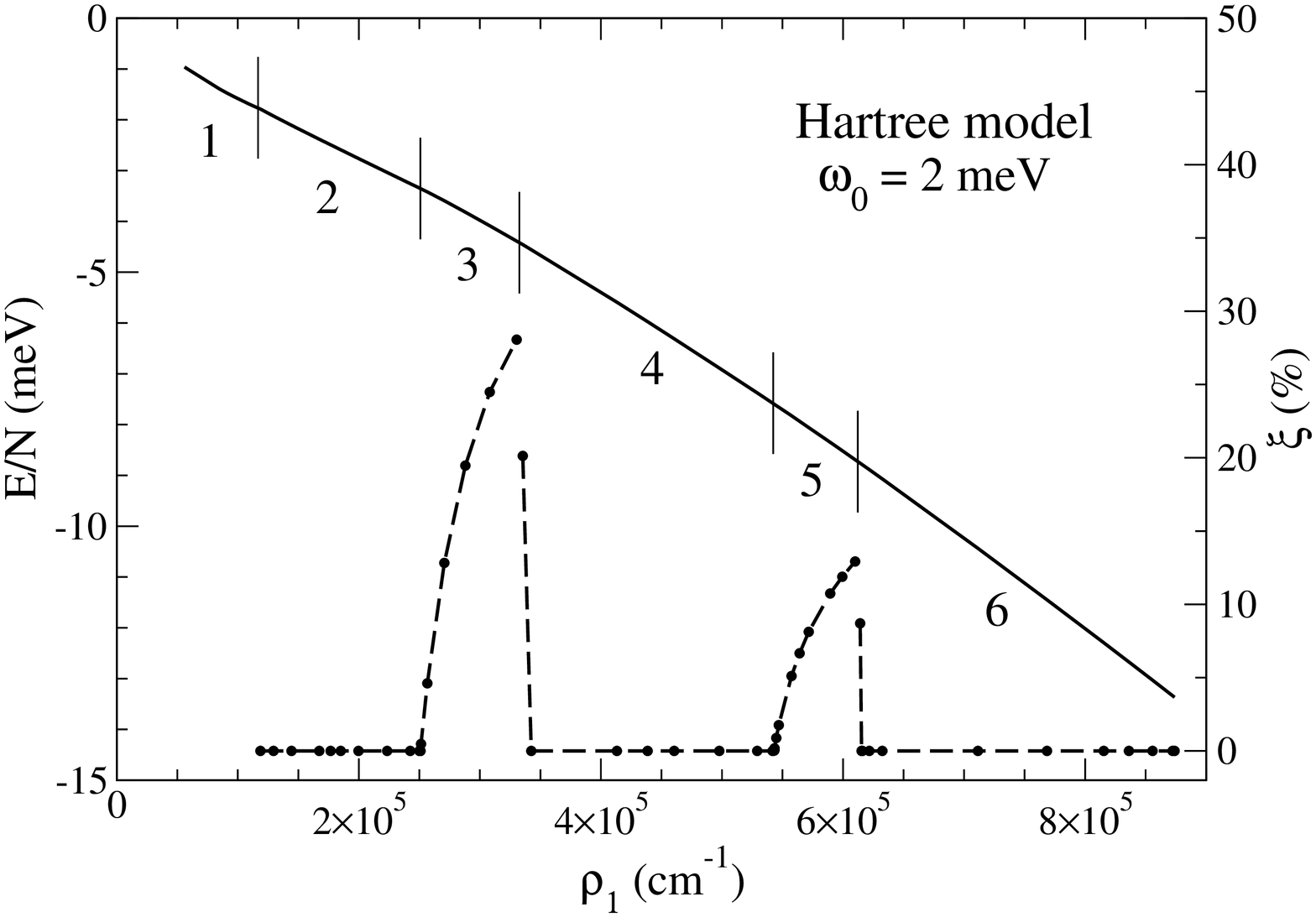}}

\caption[]{
Energy per electron (meV, left scale) and magnetization (right scale)
in the Hartree model for a parabolic confinement of $\omega_0=2$ meV
as a function of the linear density (cm$^{-1}$).
The regions separated by vertical lines correspond to the indicated
number of occupied subbands. 
For one single occupied subband, the system is fully polarized.
}
\label{fig1}
\end{figure}

\begin{figure}

\centerline{\includegraphics[height=20cm,angle=-90]{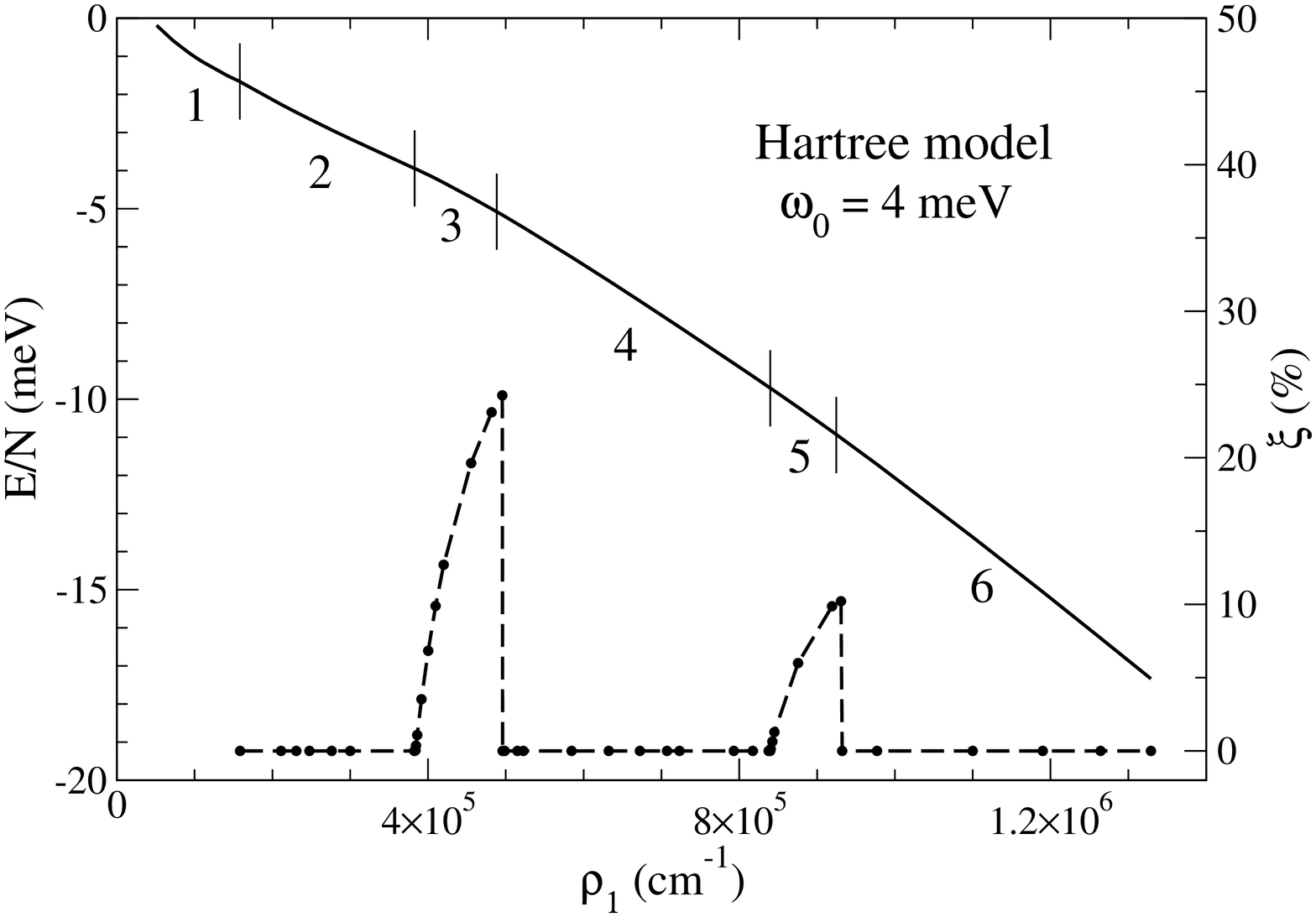}}

\caption[]{
Same as Fig. \ref{fig1} for $\omega_0=4$ meV.
}
\label{fig2}
\end{figure}

\begin{figure}

\centerline{\includegraphics[height=20cm,angle=-90]{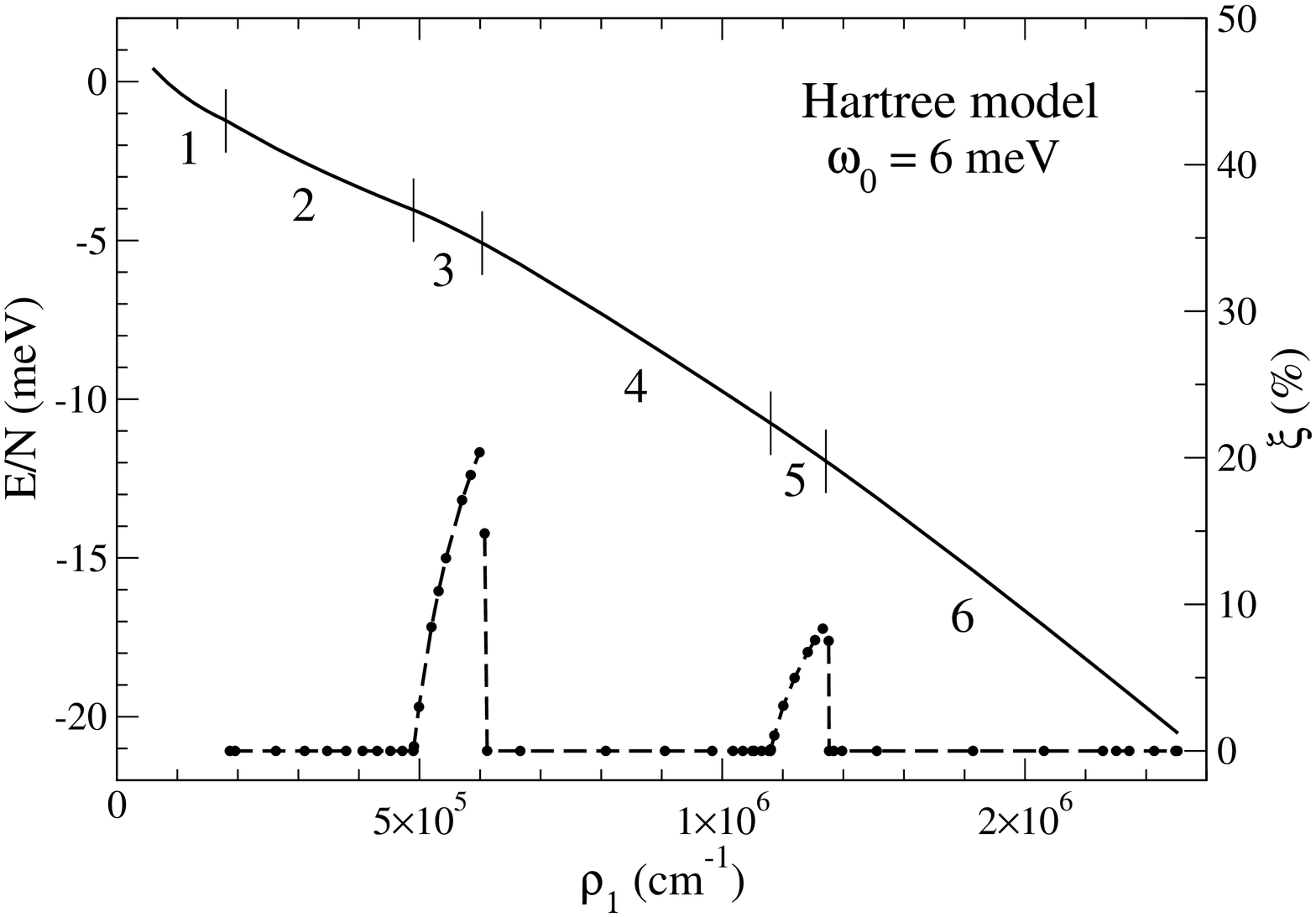}}

\caption[]{
Same as Fig. \ref{fig1} for $\omega_0=6$ meV.
}
\label{fig3}
\end{figure}

\begin{figure}

\centerline{\includegraphics[height=20cm,angle=0]{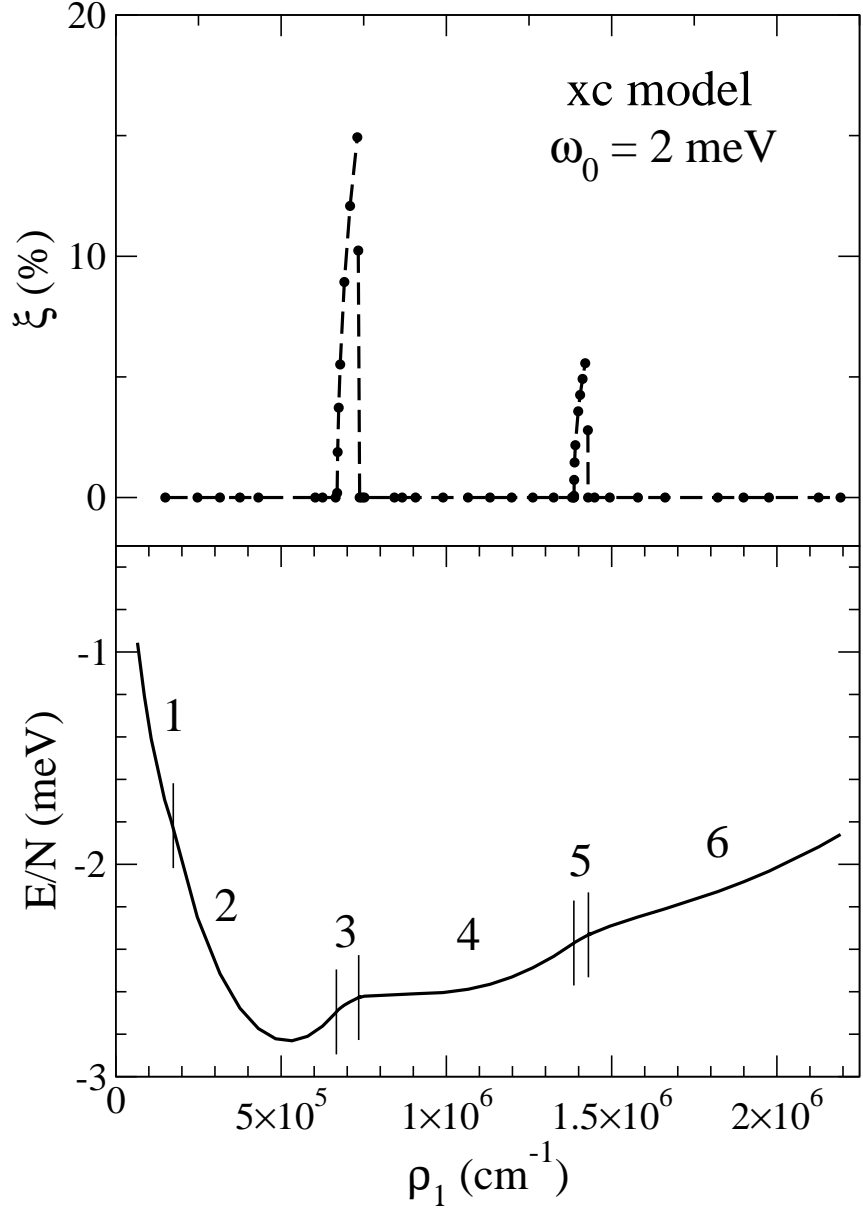}}

\caption[]{
Energy per electron (meV, bottom panel) and magnetization (top panel)
in the exchange-correlation model for a parabolic confinement of $\omega_0=2$ meV
as a function of the linear density (cm$^{-1}$).
The regions separated by vertical lines correspond to the indicated
number of occupied subbands.
For one single occupied subband, the system is fully polarized.
}
\label{fig4}
\end{figure}

\begin{figure}

\centerline{\includegraphics[height=20cm,angle=0]{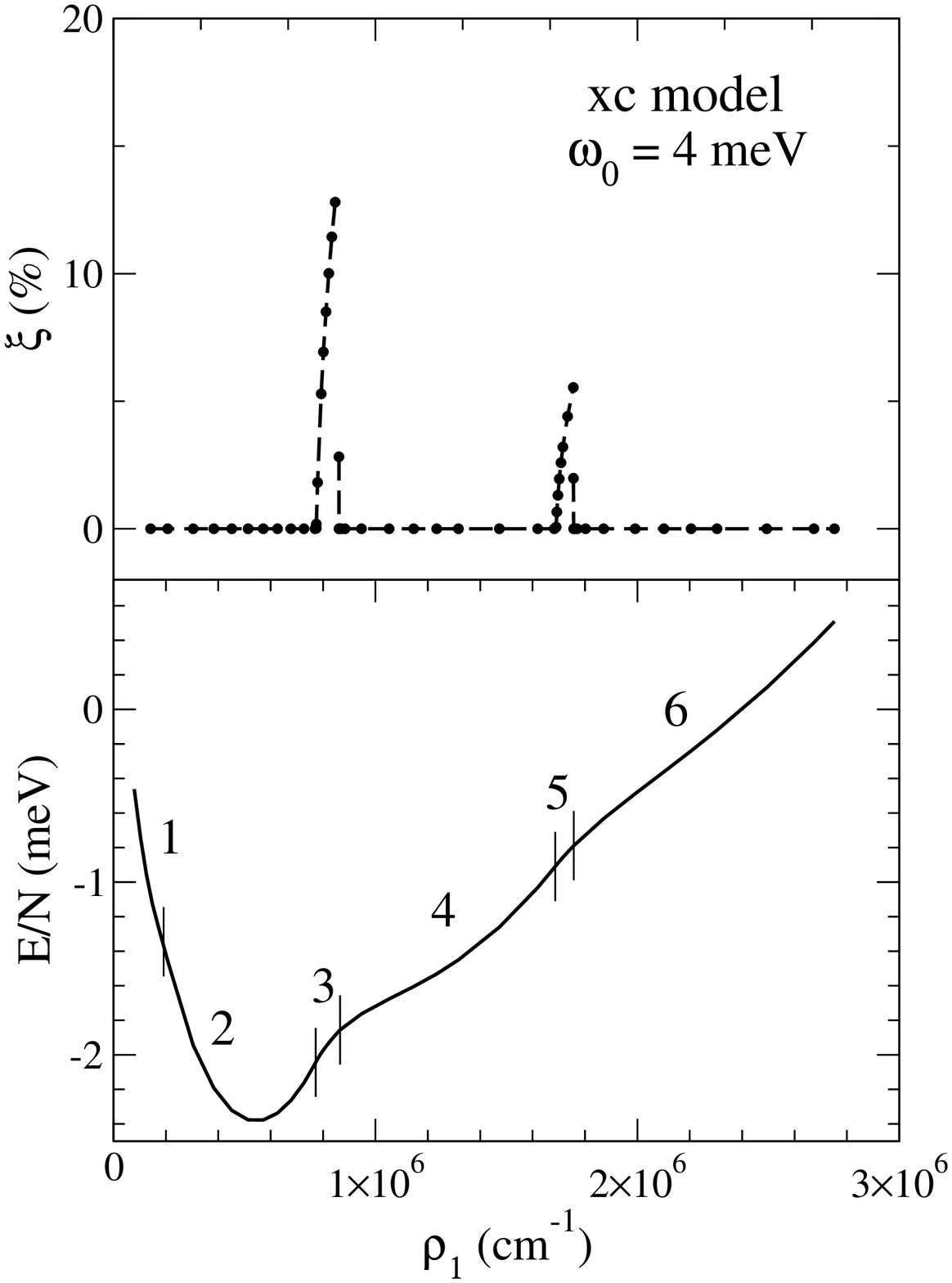}}

\caption[]{
Same as Fig. \ref{fig4} for $\omega_0=4$ meV.
}
\label{fig5}
\end{figure}

\begin{figure}

\centerline{\includegraphics[height=20cm,angle=0]{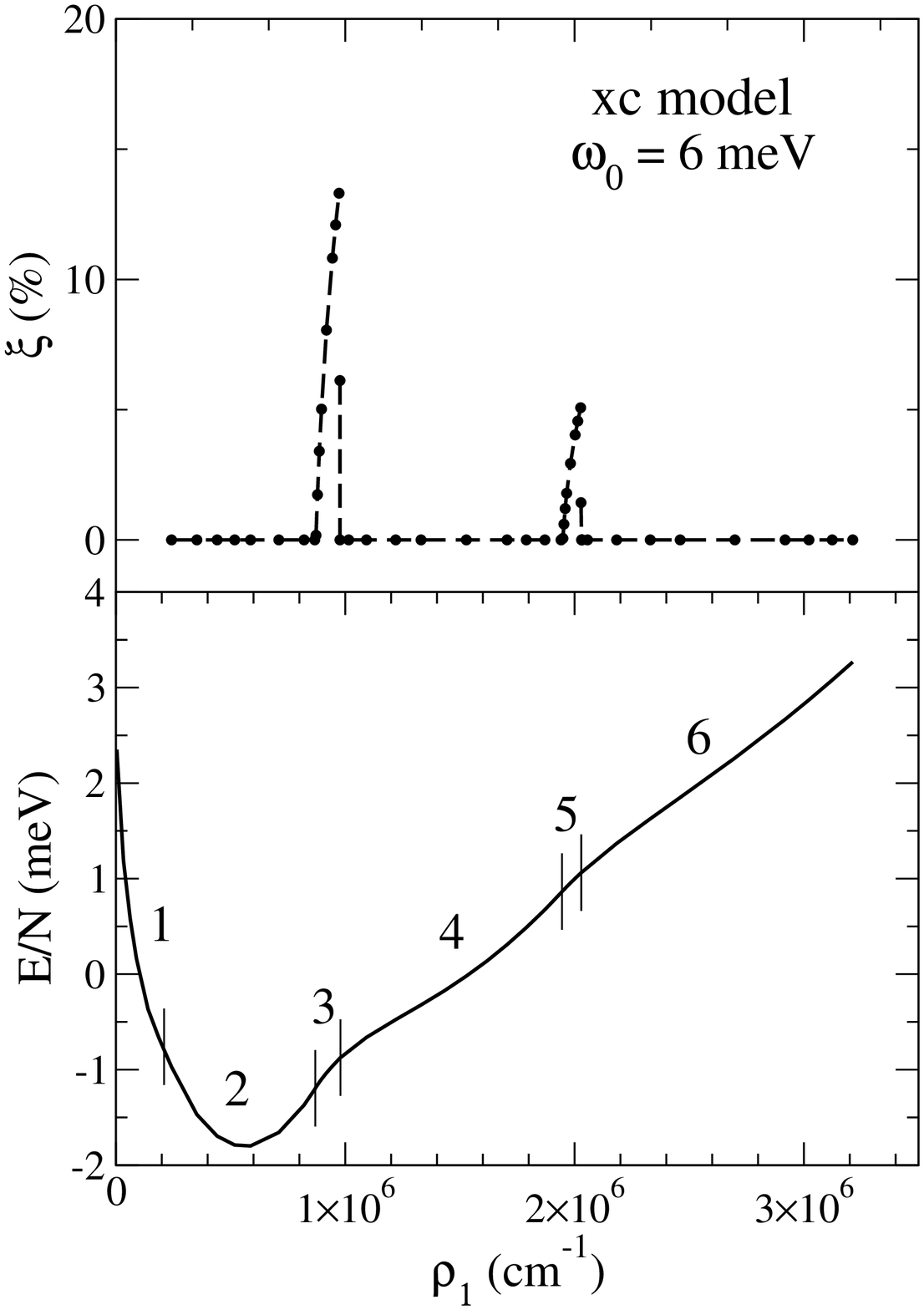}}

\caption[]{
Same as Fig. \ref{fig4} for $\omega_0=6$ meV.
}
\label{fig6}
\end{figure}

In Figs. \ref{fig1}-\ref{fig3} we have plotted $E/N$ and the magnetization 
$\xi=(\rho_1^{\uparrow}-\rho_1^{\downarrow})/\rho_1$ as a function of the one-dimensional
electron density for the Hartree model, and
in Figs. \ref{fig4}-\ref{fig6} for the exchange-correlation model. The numbers along
the $E/N$ curves correspond to the number of occupied subbands, and the vertical lines
indicate the boundary between neighbor $j$ and $j+1$ subband regions.

For both models, we have found that the transition from even-to-odd number of subbands
neighbor regions, like $2 \rightarrow3$ or $4\rightarrow5$, is smooth,
as the changes in $\xi$ indicate. On the contrary,
the transition from odd-to-even number of subbands neighbor
regions, like $1 \rightarrow2$ or $3\rightarrow4$, is abrupt, with metastability regions
-not shown in the figures- extending to the left and right of the crossing points.
Apart from the region corresponding to the first subband, which is fully polarized,
the magnetization reached in all the other odd subband regions is below the
maximum value one would naively expect, i.e.,
$1/3$ for $j=3$ and $1/5$ for $j=5$. This is due to the exchange-correlation energy, which
lifts the degeneracy of the $\uparrow,\downarrow$ subbands.
It is also worth to note that the odd subband number regions are rather narrow, 
especially
in the exchange-correlation model, getting narrower as $\rho_1$ increases.
The relevance of determining the boundaries of these density regions is that, 
in a mean field model, they fix the extension of the conductance 
plateaus, as shown in the next subsection.

\begin{figure}

\centerline{\includegraphics[height=20cm,angle=-90]{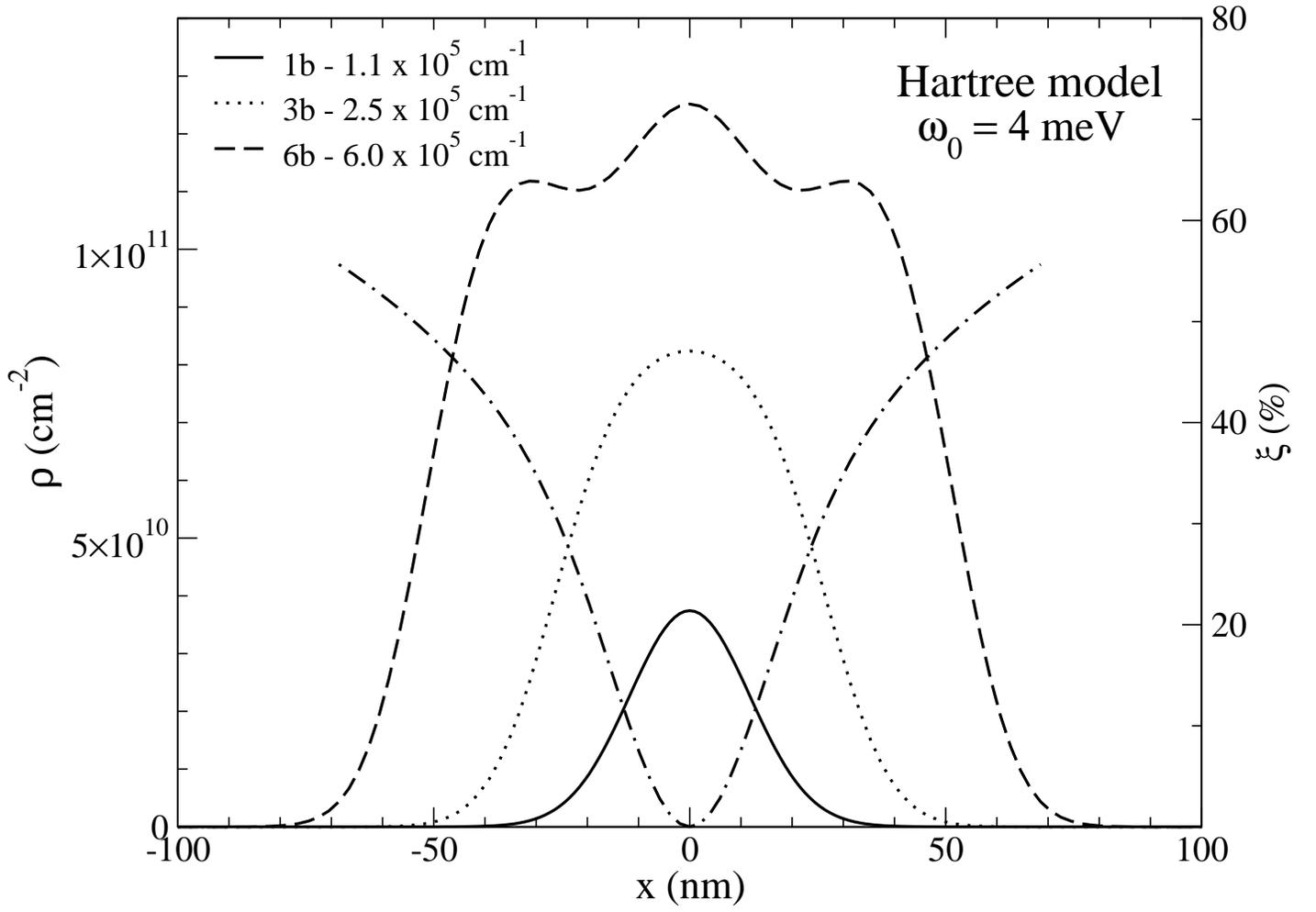}}

\caption[]{
Several density profiles of the 2D electronic density
as a function of $x$ (nm) for the Hartree  model and a  harmonic
frequency $\omega_0=4$ meV (left scale). Also shown is the local 
magnetization $\xi$ for the 3 subband case (right scale).
The values of the 1D electronic densities, which
correspond to configurations with
1, 3, and 6 occupied subbands, are indicated.
}
\label{fig7}
\end{figure}

\begin{figure}

\centerline{\includegraphics[height=20cm,angle=-90]{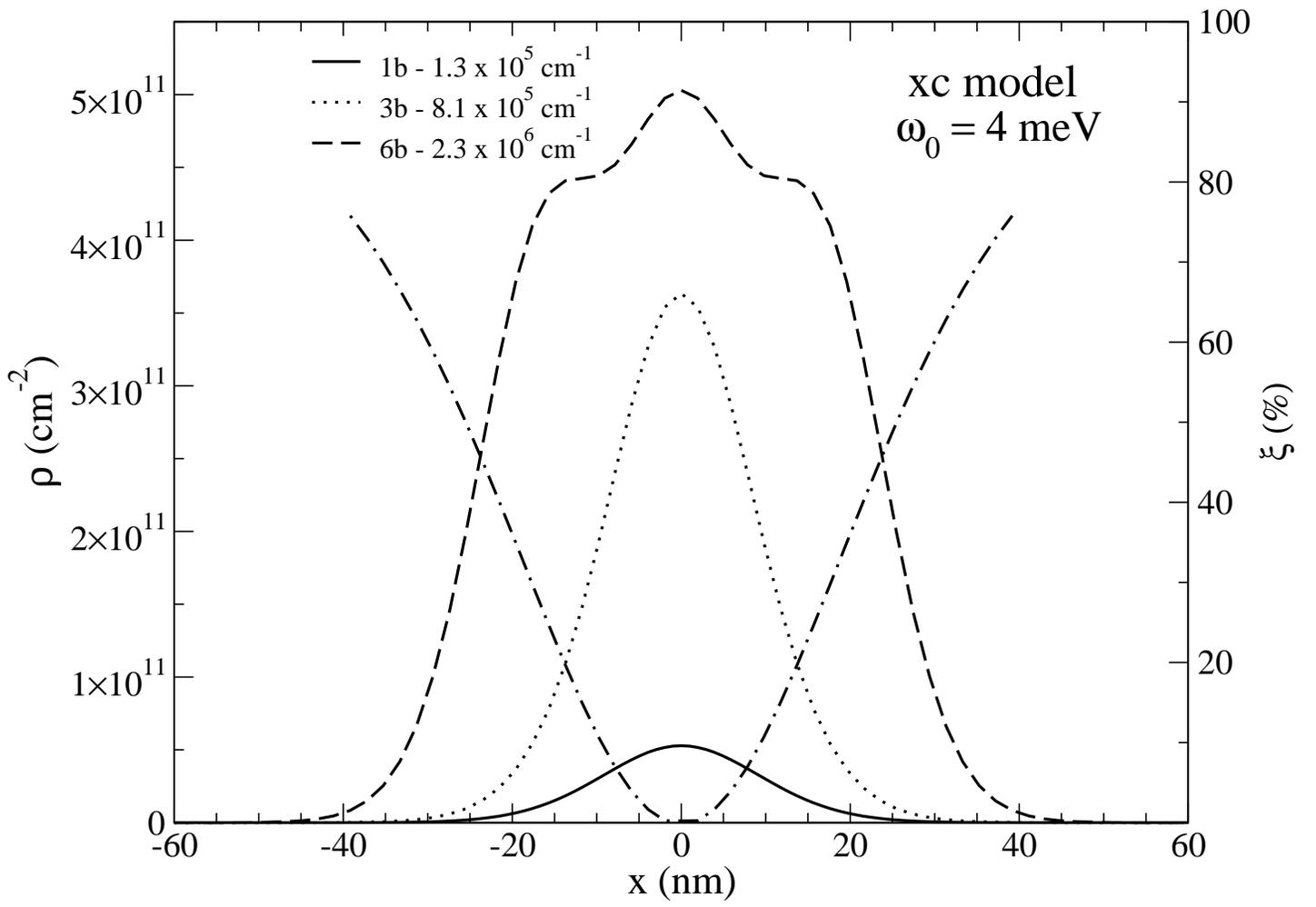}}

\caption[]{
Same as Fig. \ref{fig7} for the exchange-correlation model.
}
\label{fig8}
\end{figure}

In Figs. \ref{fig7} and \ref{fig8} we show the
density profiles of the 2D electronic  density 
for the Hartree and exchange-correlation models, respectively,
and a lateral confining potential with $\omega_0=4$ meV.
The values of the 1D electronic densities are
indicated, and correspond to configurations with 1, 3, and 6 
occupied subbands.
Also shown is the {\it local} magnetization 
$\xi= m(x)/\rho(x)$ for the 3 subband systems.
It can be seen that in this later case, the strip presents an edge magnetization that
increases as $\rho$ decreases.

Some LSDA calculations 
have found that, starting from paramagnetic two subband configurations
and decreseasing further the electronic density, for certain values
of the wire width the system may undergo a phase transition
and the ground state becomes a spin-density wave  (SDW)
\cite{Rei99,Kar05}. Using as a guide the diagram shown in Fig. 1 of
Ref. \cite{Rei99}, we have checked that, at low densities, for the
strips considered here, their LSDA ground state is
the spin polarized one. The same happens for the narrow
wires considered e.g. in Refs.
\cite{Gold96,Wan96,Ram05,Ber02,Hav04}.
We will come back to this point in the next Sec. when we discuss the 
conductance. It is worthwhile to note that the presence of a SDW 
will modify the phase diagram at the border between the one and two 
subband regions, as this band structure will disappear.

\subsection{Kohn-Sham response of the quantum strip}

The KS calculation discussed above allows also to evaluate the KS -free, in the
mean field sense- linear density
response $\chi_{0}(q,\omega)$ to a field parallel to the strip, i.e., in the $y$ direction,
which involves only intraband excitations:
\begin{eqnarray}
\label{eq12}
& &{\chi_{0}(q,\omega)\over L}=
{1\over L}\sum_{\ell}{\vert\langle \ell\vert\sum_{i=1}^N e^{i q y_i}\vert 0\rangle\vert^2
\over (\omega+ i\,\eta)^2-\omega^2_{\ell 0}}=
\nonumber\\
& &
{1\over \pi}\sum_{n,\sigma}
\int_{-\sqrt{2(\epsilon_F-\epsilon_{n,\sigma})}}^{\sqrt{2(\epsilon_F-\epsilon_{n,\sigma})}}
d k \left({1\over \omega+i\,\eta -k q -q^2/2}-
{1\over \omega+i\,\eta +k q +q^2/2}\right) \; ,
\end{eqnarray}
where $\eta$ is a small real quantity, and hence the longitudinal conductivity
associated to it \cite{Pin66}:
\begin{equation}
\label{eq13}
{\sigma(q,\omega)\over L}=i {\omega e^2\over q^2}{\chi_{0}(q,\omega)\over L}~,
\end{equation}
with the real part given by
\begin{equation}
\label{eq14}
Re{\sigma(q,\omega)\over L}=- {\omega e^2\over q^2}Im{\chi_{0}(q,\omega)\over L}~.
\end{equation}
The conductance $G$ is obtained from $Re [\sigma(q,\omega)/L]$ taking the $\omega\to0$
and $q\to0$ limits. In the small $q$ limit, for $\omega\ge0$ one gets
\begin{equation}
\label{eq15}
Im{\chi_{0}(q,\omega)\over L}=-{q\over2}\sum_{n,\sigma}
\delta(\omega-q\,k_F^{n,\sigma})~,
\end{equation}
where we have defined
\begin{equation}
\label{eq16}
k_F^{n,\sigma}=\sqrt{2(\epsilon_F-\epsilon_{n,\sigma})}~.
\end{equation}
From Eqs. (\ref{eq14}-\ref{eq15}), taking the Fourier transform, it can be easily shown that
\begin{equation}
\label{eq17}
Re{\sigma(y,\omega)\over L}= 
{ e^2\over 2\pi}\sum_{n,\sigma}\cos\left({\omega\, y\over k_F^{n,\sigma}}\right)~.
\end{equation}
Thus, in the limit $\omega\to0$ and restoring $\hbar$, we get for the conductance
\begin{equation}
\label{eq18}
G= { e^2\over h}\sum_{n,\sigma} 1 ~,
\end{equation}
where the sum runs over the occupied subbands of the strip.
One thus see that in the KS mean field model,
in the density regions we have previously determined,
the conductance takes 
the values $1/2, 1, 3/2, 2, \ldots$ in units of $2 e^2/h$.
The odd subband regions are so narrow 
that their experimental significance may be limited, especially in the more
relevant case of full screening (exchange-correlation model).
Although the KS conductance explains the gross experimental features,
it is however unable to explain fine details, like the
structures commented in the Introduction.
In the next section we 
shall try to explain these findings incorporating
interaction effects which are missing in the  KS conductance.

\section{Conductance and screened response}

We turn now our attention to
the conductivity of a quantum strip as the screened response
to the sum of the applied external plus the local induced field. 
For simplicity, we address only the lowest density cases, namely when
the quantum strip is ferromagnetic and only one
subband is occupied, and when  the strip
is paramagnetic and only two subbands (spin-up and spin-down)
are occupied, being degenerate and equally populated. These two situations
are the more interesting ones from an experimental viewpoint.
The generalization of the method to the case with more than 
two subbands is straightforward but requires a more elaborate
calculation.

The calculation of the conductance from the screened response is out 
of any ambiguity when the screened response corresponds to a 
`fictious' neutral system with
the same interaction law as the screened electron-electron Coulomb
interaction \cite{Pin66}. This can be fully achieved for 
the
exchange-correlation model previously discussed. In the case of the 
Hartree model, where local non neutrality is assumed, the calculation of 
the screened response is less clear and strongly model dependent. 
For this reason, in the following we develop the theory only for 
the exchange-correlation model and just make a few comments on the 
Hartree model results at the end of this Sec.

Our starting point is the relation \cite{Izu61,Pin66}
\begin{equation}
\label{eq19}
{\sigma(q,\omega)\over L}=i {\omega e^2\over q^2}{\chi_{sc}(q,\omega)\over L}~,
\end{equation}
where  $\chi_{sc}(q,\omega)$ is the screened response function. 
It is related to the linear density response function $\chi(q,\omega)$ by\cite{Pin66}
\begin{equation}
\label{eq20}
\chi(q,\omega)={\chi_{sc}(q,\omega)\over 1-v(q)\chi_{sc}(q,\omega)}~,
\end{equation}
where
\begin{equation}
\label{eq21}
v(q)=2\int d x \,d x^{\prime}|\phi_0(x)|^2\,|\phi_0(x^{\prime})|^2 K_0[q(x-x^{\prime})]
\end{equation}
is the quasi-1D Fourier transform of the Coulomb potential (see the
Appendix),
$\phi_0(x)=\phi_0^{\uparrow}(x)$ for the ferromagnetic case, and
$\phi_0(x)=\phi_0^{\uparrow}(x)=\phi_0^{\downarrow}(x)$ for the paramagnetic case.
In the RPA, the electron response to the total field (the sum of the external
field plus the local induced field), $\chi_{sc}(q,\omega)$, is approximated by the 
free response
\begin{equation}
\label{eq22}
\chi_{sc}(q,\omega)=\chi_{0}(q,\omega)~, 
\end{equation}
yielding
\begin{equation}
\label{eq222}
\chi^{RPA}(q,\omega)={\chi_{0}(q,\omega)\over 1-v(q)\chi_{0}(q,\omega)}~.
\end{equation}
To go beyond the RPA, the short-range
electron-electron correlations may be taken into account by modifying Eq. (\ref{eq22})
as
\begin{equation}
\label{eq23}
\chi_{sc}(q,\omega)={\chi_{0}(q,\omega)\over 1+{v(q)\over L}
{\cal G}(q,\omega)\chi_{0}(q,\omega)}~,
\end{equation}
where the dynamic local field correction ${\cal G}(q,\omega)$
has been introduced \cite{Iwa84,Lip04}. Eq.
(\ref{eq23}) is  the most general way to express $\chi_{sc}(q,\omega)$ in terms of 
 the free response function $\chi_{0}(q,\omega)$, and yields for $\chi(q,\omega)$
\begin{equation}
\label{eq233}
\chi(q,\omega)=
{\chi_{0}(q,\omega)\over 1-{v(q)\over L} 
[1-{\cal G}(q,\omega)] \chi_{0}(q,\omega)}~.
\end{equation}
In the following we are only interested in
the $\omega=0$ limit, so the frequency dependence of the local field correction
is suppressed   
\begin{equation}
\label{eq24}
\chi_{sc}(q,\omega)={\chi_{0}(q,\omega)\over 1+{v(q)\over L}
\,{\cal G}(q)\,\chi_{0}(q,\omega)}~.
\end{equation}
In this form, the frequency dependence of $\chi_{sc}(q,\omega)$ comes only from $\chi_{0}(q,\omega)$. 
An important property of the local field correction   is \cite{Iwa84}
\begin{equation}
\label{eq25}
\lim_{q\to0} v(q)\,{\cal G}(q)=v(0)\,{\cal G}(0)=
{1\over \rho_1^2 K_0 }\left(1-\frac{K_0}{K}\right)~,
\end{equation}
where $K$ is  the compressibility of the system and $K_0$ its free value. 
In the situations we are considering, we have
$K_0=1/(\pi^2 \rho_1^3)$ for
the ferromagnetic case, and $K_0=4/(\pi^2 \rho_1^3)$ for the paramagnetic case,
and the compressibility $K$ can be calculated from the standard -thermodynamical-
expression
\begin{equation}
\label{eq26}
{1\over K}= \rho_1^2\left(\rho_1{\partial^2 E/N\over\partial\rho_1^2}
+2{\partial E/N\over\partial\rho_1}\right)~. 
\end{equation}
For the locally neutral system the energy functional contains only the kinetic and
exchange-correlation pieces plus the confining potential in the $x$-direction.
One gets
\begin{equation}
\label{eq27}
E/N={\pi^2\over c}\rho_1^2  + 
{1\over\rho_1}\int d x \,{\cal E}_{xc}(x) +const. ~,
\end{equation}
where  $c=6$  for the ferromagnetic, and $c=24$ for the paramagnetic case.
Hence
\begin{equation}
\label{eq28}
{1\over K}={1\over K_0}+\rho_1^2 I~,
\end{equation}
where we have defined 
\begin{equation}
\label{eq288}
I =  \int d x \,{\partial^2 {\cal E}_{xc}\over\partial\rho^2}\,
\vert\phi_0(x)\vert^4~.
\end{equation}

From Eqs. (\ref{eq12}) and (\ref{eq24}), we write
in the small $q$ limit
\begin{equation}
\label{eq29}
{L\over\chi_{sc}(q,\omega)}={L\over\chi_{0}(q,\omega)}+v(0){\cal G}(0)=
{\omega^2-(k_F^0 q)^2\over c^{\prime} k_F^0 q^2/\pi}+v(0){\cal G}(0)~,
\end{equation}
where $k_F^0=\sqrt{2(\epsilon_F-\epsilon_0)}$ with $\epsilon_0=\epsilon_0^{\uparrow}$
and  $c^{\prime}=1$ for the ferromagnetic case, and with
$\epsilon_0=\epsilon_0^{\uparrow}=\epsilon_0^{\downarrow}$ and $c^{\prime}=2$ 
for the paramagnetic case. Hence
\begin{equation}
\label{eq30}
{1\over L}\,\chi_{sc}(q,\omega)=
\gamma c^{\prime}{k_F^0 q^2\over\pi\gamma}{1\over \omega^2-(k_F^0 q/\gamma)^2}
=\gamma\,{1\over L}\,\chi_0(q,\omega,k_F^0/\gamma)~,
\end{equation}
where 
\begin{equation}
\gamma=
\left(1-c^{\prime}{v(0){\cal G}(0)\over\pi k_F^0}\right)^{-1/2}=
\sqrt{K\over K_0}~.
\label{eq31}
\end{equation}
It is now possible to calculate the real part of the conductivity and thus
the conductance along the line of Eqs. (\ref{eq12}-\ref{eq18}),
using $\chi_{sc}(q,\omega)/L=\gamma\,\chi_0(q,\omega,k_F^0/\gamma)/L$
at the place of $\chi_0(q,\omega,k_F^0)/L$. One easily gets 
\begin{equation}
\label{eq32}
G= { e^2\over h}\sqrt{K\over K_0}
\end{equation}
for the ferromagnetic case, and
\begin{equation}
\label{eq33}
G= {2 e^2\over h}\sqrt{K\over K_0}
\end{equation}
for the paramagnetic case.
In both situations the ratio $K/K_0$ is calculated from the expression
\begin{equation}
\label{eq34}
{K\over K_0}={1\over 1+{2c^{\prime}\over \pi^2\rho_1} I} \; ,
\end{equation}
where $I$ is defined in Eq. (\ref{eq288}) with
$\phi_0(x)=\phi^{\uparrow}_0(x)$ for the ferromagnetic case, and
 $\phi_0(x)=\phi^{\uparrow}_0(x)=\phi^{\downarrow}_0(x)$ for the paramagnetic case.
Eqs. (\ref{eq31}-\ref{eq34}) can be also obtained calculating directly
the screened response function in the time-dependent local 
spin-density approximation  (TDLSDA),
as done in the Appendix. The above derivation is however more general and
shows that Eqs. (\ref{eq32}-\ref{eq33}) are model independent, although the
way in which we have calculated  $K$ is clearly model dependent.

\begin{figure}

\centerline{\includegraphics[height=20cm,angle=-90]{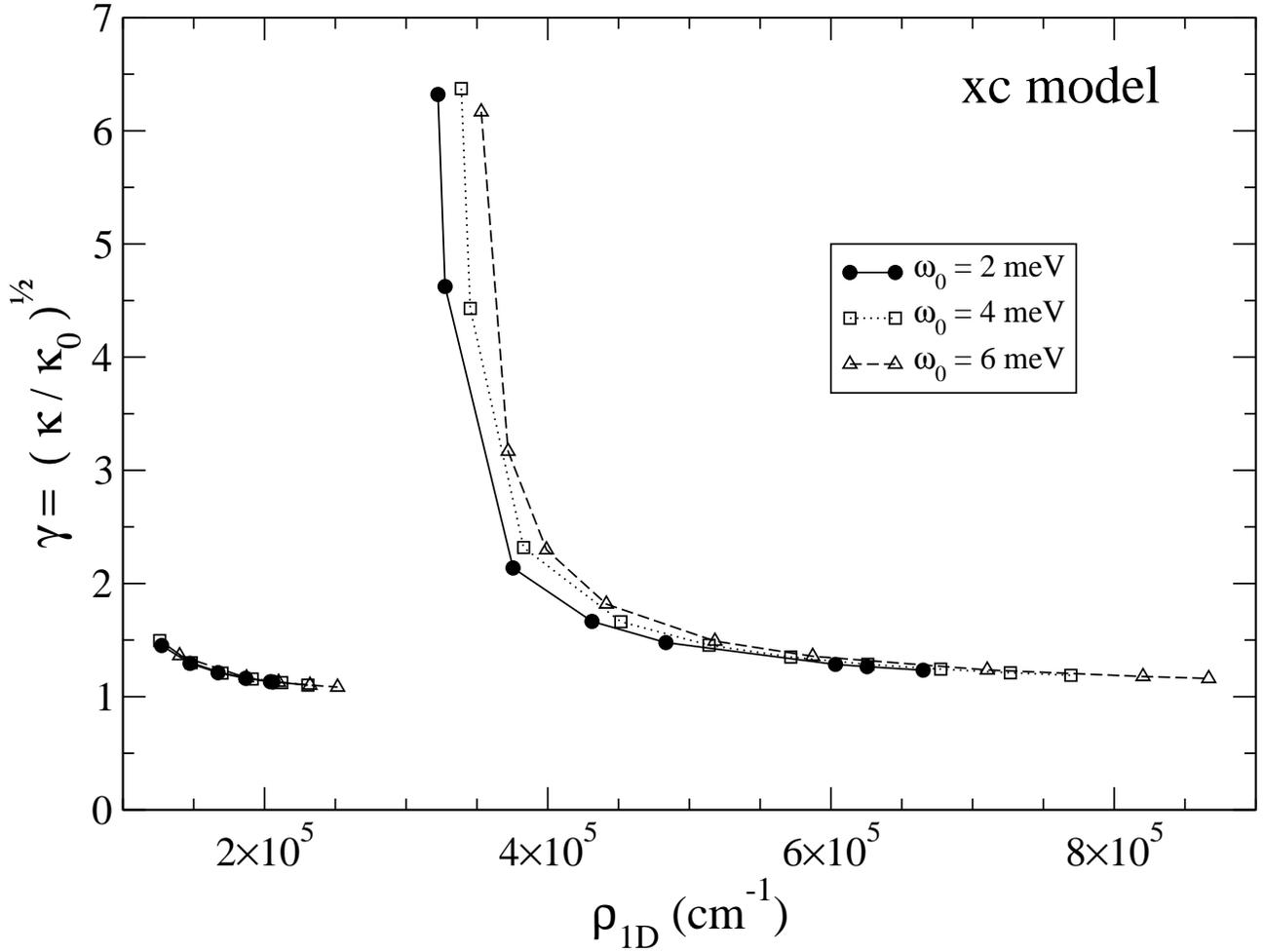}}

\caption[]{
Ratio $\gamma= \sqrt{K/K_0}$ as a function of the linear density $\rho_1$ 
(cm$^{-1}$) for the exchange-correlation model and three values of the
harmonic frequency $\omega_0$. The lines have been drawn
to guide the eye.
}
\label{fig9}
\end{figure}

The ratio $\gamma=\sqrt{K/K_0}$  is plotted
in Fig. \ref{fig9} as a function of
$\rho_1$ for different values of the frequency of the lateral confining potential±
$\omega_0$. One sees from this figure that when the strip is in
the first -polarized- subband, $\gamma$ approximately
ranges from 1.5 to 1 for all the confinements considered here, yielding for this subband
a conductance which goes from $G\simeq 0.7 (2 e^2/h)$ to $G\simeq 0.5 (2 e^2/h)$  
with increasing density, in agreement with the experimental data of Reilly 
et al. \cite{Rei01}. The discontinuity
of $\gamma$ and $K$ when passing from one  to two
subbands reflects the phase transition occurring in the system. 
After that, the paramagnetic state has a conductance that, starting from values
a slightly larger than  $2 e^2/h$, decreases with increasing density to the 
measured value $G\simeq  2 e^2/h$. Despite that our model for an infinite
quantum strip is obviously an oversimplification of the actual experimental device,
it yields a qualitative agreement with measurements, especially  the observed
density dependence of the conductance.
We want to comment that at values of $\rho_1$ lower than the ones
considered here,
$\gamma$, as calculated from Eqs. (\ref{eq31}-\ref{eq34}),  becomes imaginary 
because the compressibility is negative. 
This result for the compressibility is well known \cite{Cam96}, and it 
appears also for the electron gas in two and three  dimensions, reflecting
the failure at low densities, of the jellium model  in which  the
background of positive ions is kept indefinitely rigid.

The phase transition between one and two bands makes room
to the appearance of a SDW in the system, which we have not
found in the calculations because of the way we have carried
them out in practice. The instability of the paramagnetic
configuration, as manifested by the divergency of $K/K_0$, shows
that the ground state of the strip must be different from the
paramagnetic one, and cannot yet be the ferromagnetic phase, which
has a higher energy and only becomes the ground state at lower
densities. Indeed, we have checked that the value of the
ratio $C_F = \hbar \omega_0 /\epsilon_F$, where $\epsilon_F$ is
the Fermi energy of the 1D electron gas at the given density,
introduced in Ref. \cite{Rei99} to characterize the one-dimensionality
of the system, takes values that are compatible with the presence of
a SDW, as shown in Fig. 1 of that reference.
To ascertain the role that a SDW might play in the conductance of a
quantum strip is beyond the present model, which assumes the existence
of a band structure from the start.

Finally, we discuss the effect that the non local charge neutrality
implicit in the Hartree model of Sect. II has on the value of $\gamma$.
A straightforward generalization of the calculation developed
in the Appendix for the exchange-correlation model shows that, for
the logarithmic Hartree potential model (\ref{eq10}),
one just has to change everywhere
\begin{equation}
\label{eq289}
I \rightarrow 
I -2 \int dx\, dx^{\prime}\,\vert\phi_0(x)\vert^2\,\vert\phi_0(x^{\prime})\vert^2
\,ln\left|
{x-x^{\prime}\over a}\right| \, .
\end{equation}
Differently from the exchange-correlation contribution (\ref{eq288}),
which is attractive,
the above Hartree contribution is repulsive and may change the value of $\gamma$
in a relevant way. 
In the case of two occupied subbands, the effect of the Hartree
contribution is to make the ratio $K/K_0$ smaller than one, and so
the value of $\gamma$. 
Quantitatively, in the density range where the ground state of the system
is paramagnetic, the value of
$\gamma$ varies from 1 to 0.9, to be compared with the values slightly
larger than 1 obtained with the assumption of charge neutrality.
Both models yield values of $G$ which are  close
to the measured value $G\simeq  2 e^2/h$. 
We want to stress again that the results obtained when one adds a term which
causes that charge neutrality is locally violated are very model
dependent.

\section{CONCLUSIONS}
Within LSDA,
we have studied the ground state structure of quantum strips for a
range of electronic densities and several
strenghts of the lateral confining potential.
As the 1D electron density increases, the system can be
in different phases characterized by different values of the
the magnetization. These phases correspond to the filling of an increasing
number of electronic subbands. Due to the exchange-correlation interaction,
we have found that for odd numbers of occupied subbands larger than one,
the system adquires some edge magnetization. The width of the non-paramagnetic
density regions is rather narrow, and it decreases as the density
increases.

We have employed two approximations for the electron-electron
interaction commonly employed to address the structure of
quantum wires. The first one is a logarithmic Hartree potential
successfully
used in the past to address the infrared response of quantum wires
in the RPA \cite{Gud95,Ste96}. The second one supposes that the
direct electron-electron interaction is fully screened by a jellium
background, and the system is neutral at a local scale.

We have used this second approximation to address the conductivity of quantum
strips when one or two subbands are occupied in the ground state,
going beyond the mean field or random-phase approximations. We have
found, in a model independent way, that the physical
parameter that determines the value of conductance is
the ratio $K/K_0$ of the compressibility of the system over the free one. 
This result has been used to obtain the conductance $G$ of the system
in the LSDA. We have found that when only one subband is occupied,
$G$  takes a value  close to 0.7(2e$^2/h$) and decreases as the
electron density increases, in  agreement with experiments.
When two subbands are occupied and the system becomes paramagnetic, $G$
takes a value near (2e$^2/h$), showing a similar decreasing behaviour
with increasing electron density as for one subband.

\appendix*
\section{}

In this Appendix we calculate the TDLSDA density response function
of a quantum strip for the two cases -ferromagnetic and paramagnetic-
discussed in Sec. III,
and derive the screened response in the same approximation.
We start from the time-dependent KS equations
\cite{Lip04} in an external, time oscillating
field $\lambda\,(\hat{O}^{\dagger}e^{-i\omega t}+\hat{O} e^{i\omega t})$ with
$\hat{O}=\sum ^{N}_{i=1}e^{-i q y_{i}}$, in the $y$-direction along the wire:
\begin{eqnarray}
i\frac{\partial }{\partial t}\varphi\left( x,y,t\right)
&=&\left\{-\frac{1}{2}\nabla_x^{2}-\frac{1}{2} \nabla_y^{2}+
\frac{1}{2}\omega_0 \,x^2 +
\int d x^{\prime}\,d y^{\prime}\frac{\rho(x^{\prime},y^{\prime},t)-
\rho_J}{\sqrt{(x-x^{\prime})^2+(y-y^{\prime})^2}}
\right.
\nonumber
\\
&&\left. +v_{xc}[\rho(x,y,t)]
+\lambda \, [e^{i\left( q y-\omega
t\right) }
+e^{-i\left( q y-\omega t\right) }] \right\}
\varphi\left(x,y,t\right) \, ,
\label{eq A1}
\end{eqnarray}
where $\rho_J$ is the jellium density.
For the ferromagnetic case $\varphi(x,y)$
is the single particle wave function of the electrons
in the lowest fully polarized subband 
and $\rho=\rho^{\uparrow}$.
In the
paramagnetic case in which electrons fill the two lowest degenerate spin-up
and down subbands,
$\varphi(x,y)=\varphi^{\uparrow}(x,y)=\varphi^{\downarrow}(x,y)$ 
and $\rho=\rho^{\uparrow}+\rho^{\downarrow}$.
Since the time oscillating field is in the $y$-direction, electron density oscillations
are induced by the field only along this direction. We write
\begin{equation}
\label{eq A2}
\rho \left(x,y,t\right)= \rho_0 + \delta \rho \left(x,y,t\right) \; ,
\end{equation}
where $\rho_0$ is the density of the unperturbed ground state
which we suppose to be equal to the jellium density for the neutral system
we are considering in this Appendix, and
\begin{equation}
\label{eq A3}
\delta \rho \left(x,y,t\right)=\vert\phi_0(x)\vert^2\delta\rho(y,t)
\end{equation}
with
\begin{equation}
\delta \rho \left(y,t\right)=\delta\rho\left( e^{i\left(q 
y-\omega t\right) }
+e^{-i\left(q y-\omega t\right) }\right)~,
\label{eq A4}
\end{equation}
$\delta\rho$ being a  constant to be determined.
The density fluctuation Eqs. (\ref{eq A2})-(\ref{eq A4}) induce fluctuations 
in the density operator  $F=\sum ^{N}_{i=1}e^{i q y_{i}}$ given by
\begin{equation}
\delta F(\hat{O},\omega)=\langle\psi(t)|F|\psi(t)\rangle
-\langle 0|F|0\rangle=
\int d x \,d y \,e^{i q y}[ \rho(x,y,t)
-\rho_{0}]_{\hat{O}}= L \,e^{i\omega t}\delta\rho~,
\label{eq A5}
\end{equation}
where $L$ is the length of the strip. Moreover, the
dynamic polarizability  is given by
\begin{equation}
\chi(q,\omega)=\frac{L\delta\rho }{\lambda } ~.
\label{eq A6}
\end{equation}
To determine $\delta\rho$
we now use  that the wave function $\varphi(x,y,t)$ can be factorized
in a part depending only on $x$, $\phi_0(x)$, which is not affected by the external field,
and a part which depends on $y$ and time due to the external field, $\varphi(y,t)$.
Multiplying Eq. (\ref{eq A1}) on the left by
 $\phi_0^{*}(x)$ and integrating over $x$  we get
\begin{eqnarray}
i\frac{\partial }{\partial t}\varphi\left(y,t\right)
&=&\left\{-\frac{1}{2}\nabla_y^{2} + const.
+\int d x \,d x^{\prime}\,d y^{\prime}\,\vert\phi_0(x)\vert^2\frac{\rho(
x^{\prime},y^{\prime},t)-\rho_J}{\sqrt{(x-x^{\prime})^2+(y-y^{\prime})^2}}
\right.
\nonumber
\\
&&\left. +\int d x \,v_{xc}[\rho(x,y,t)]\,\vert\phi_0(x)\vert^2 +
\lambda\, [ e^{i\left( q y-\omega t\right) }
+e^{-i\left( q y-\omega t\right) }] \right\}
\varphi\left(y,t\right) \, .
\label{eq A7}
\end{eqnarray}
We then insert $\rho(x,y,t)$ of Eqs. (\ref{eq A2})-(\ref{eq A4}) into
(\ref{eq A7})
and linearize the equations. This means writing the self-consistent
KS  mean
field entering Eq. (\ref{eq A7}), i.e.
$V_{KS}[x,y,\rho(x,y,t)]=V_H[x,y,\rho(x,y,t)]+v_{xc}[x,y,\rho(x,y,t)]$, as
\begin{equation}
V_{KS}[x,y,\rho(x,y,t)]=V_{KS}(x,y,\rho_0)
+\left.\frac{\partial V_{KS}}{\partial\rho(x,y,t)}\right|_{\rho=\rho_0}\,
\delta \rho(x,y,t) ~.
\label{eq A8}
\end{equation}
Therefore, from Eqs. (\ref{eq A7}) and (\ref{eq A8}) we obtain:
\begin{equation}
i\frac{\partial }{\partial t}\varphi\left(y,t\right)=
\left\{-\frac{1}{2}\nabla_y^{2} + const.
+[(v(q) +I)\delta\rho+\lambda]\, \left( e^{i\left( q y-\omega
t\right) }
+e^{-i\left( q y-\omega t\right) }\right) \right\}
\varphi\left(y,t\right) \, ,
\label{eq A9}
\end{equation}
where $v(q)$ is the Fourier transform of the Hartree potential
\begin{equation}
\label{eq A10}
v(q)=2\int d x \,d x^{\prime}\vert\phi_0(x)\vert^2\vert\phi_0(x^{\prime})\vert^2 
K_0[q(x-x^{\prime})]~,
\end{equation}
and $I$ is given by
\begin{equation}
\label{eq A11}
I=  \int d x \,{\partial v_{xc}\over\partial\rho}\,\vert\phi_0(x)\vert^4 ~.
\end{equation}
This quantity represents the exchange-correlation contribution to the
residual interaction. Neglecting this term in Eq. (\ref{eq A9}) yields 
the RPA which has been used to describe collective
excitations of quantum wires by several authors 
\cite{Gud95,Ste96,Das85,Gol90,Ago98}.

Eq. (\ref{eq A9}) can be rewritten as
\begin{equation}
i\frac{\partial }{\partial t}\varphi\left(y,t\right)=
\left\{-\frac{1}{2}\nabla_y^{2} + const.
+\lambda^{\prime}\, [ e^{i\left( q y-\omega
t\right) }
+e^{-i\left( q y-\omega t\right) }] \right\}
\varphi\left(y,t\right) 
\label{eq A12}
\end{equation}
with
\begin{equation}
\lambda^{\prime}(q)=\lambda+[v(q)+I]\delta\rho \, .
\label{eq A13}
\end{equation}
Equation
(\ref{eq A12}) coincides with that of a non-interacting system coupled to an
external time oscillating field, with a coupling constant $\lambda
^{\prime }$
given by Eq. (\ref{eq A13}). For such a system, the density response
function is
the -single particle- free response $\chi_0(q,\omega)$  we
have studied in Sec. II.   From Eq. (\ref{eq A6})
and from
the analogous relation for the free response function
\begin{equation}
\chi_0(q,\omega)=\frac{L \delta \rho }{\lambda ^{\prime }(q) } ~,
\label{eq A14}
\end{equation}
we obtain
\begin{equation}
\lambda\,\chi(q,\omega)=
\lambda ^{\prime }(q)\,\chi_0(q,\omega)=L \delta\rho~.
\label{eq A15}
\end{equation}
The solution of these equations is 
the TDLSDA response function
\begin{equation}
\chi^{TDLDA}(q,\omega)={\chi_0(q,\omega)\over 1 -{v(q)+I\over L}
\chi_0(q,\omega)} ~.
\label{eq A16}
\end{equation}
The RPA response function is obtained just neglecting the interaction term 
$I$ in the above equations:  
\begin{equation}
\chi^{RPA}(q,\omega)={\chi_0(q,\omega)\over 1 -{v(q)\over L}
\chi_0(q,\omega)} ~.
\label{eq A17}
\end{equation}
In the RPA $\lambda ^{\prime }=\lambda+v(q)\,\delta\rho$, and electrons respond
like non-interacting particles to the combined effect
of the external field 
$\lambda\,(\hat{O}^{\dagger}e^{-i\omega t}+\hat{O} e^{i\omega t})$
plus the Coulomb local polarization field $v(q)\,\delta\rho$
induced by the density fluctuation. 
Since the screened response is by definition the linear response 
of the system to the sum of the external field  plus the Coulomb 
polarization field 
\begin{equation}
\lambda\,\chi(q,\omega)=
\left[\lambda+\delta\rho v(q)\right]\chi_{sc}(q,\omega)=L \,\delta\rho 
\label{eq A19}
\end{equation}
\begin{equation}
\chi(q,\omega)={\chi_{sc}(q,\omega)\over 1 -{v(q)\over L}\chi_{sc}(q,\omega)} ~,
\label{eq A20}
\end{equation}
one sees immediately that in the RPA
\begin{equation}
\chi^{RPA}_{sc}(q,\omega)=\chi_0(q,\omega) ~,
\label{eq A21}
\end{equation}
whereas in the TDLDA one gets
\begin{equation}
\chi^{TDLDA}_{sc}(q,\omega)={\chi_0(q,\omega)\over 1 -{I\over L}
\chi_0(q,\omega)} ~.
\label{eq A22}
\end{equation}
The response functions described above refer to 
electrons in the lowest  subband,
$\chi(q,\omega)=\chi^{\uparrow}(q,\omega)$, 
or filling the two lowest  spin-up
and -down degenerate subbands,
$\chi(q,\omega)=\chi^{\uparrow}(q,\omega)+\chi^{\downarrow}(q,\omega)$.
The formalism can be
easily generalized to the case in which more than two subbands are
filled. 


\section*{ACKNOWLEDGMENTS}
E. L. would like to thank HPC Europa, node of Barcelona,
contract R113-CT-2003-506079, for financial support.
This work has been supported in part by DGI, Spain,
Grant FIS2005-01414.

\vfill\eject
\newpage

\end{document}